# Rethinking CO Antibiosignatures in the Search for Life Beyond the Solar System

Edward W. Schwieterman[1,2,3,4], Christopher T. Reinhard[3,5], Stephanie L. Olson[3,6], Kazumi Ozaki[3,5,7], Chester E. Harman[3,8,9], Peng K. Hong[10], and Timothy W. Lyons[1,3]

[1] Department of Earth Sciences, University of California, Riverside, CA, USA
[2] NASA Postdoctoral Program, Universities Space Research Association, Columbia, MD, USA
[3] NASA Astrobiology Institute, Alternative Earths & Virtual Planetary Laboratory Teams, USA
[4] Blue Marble Space Institute of Science, Seattle, WA, USA
[5] School of Earth and Atmospheric Sciences, Georgia Institute of Technology, Atlanta, GA, USA
[6] Department of Geophysical Sciences, University of Chicago, Chicago, IL, USA
[7] Department of Environmental Science, Toho University, Tokyo, Japan
[8] NASA Goddard Institute for Space Studies, New York, NY, USA
[9] Department of Applied Mathematics and Applied Physics, Columbia University, New York, NY, USA
[10] Planetary Exploration Research Center, Chiba Institute of Technology, 2-17-1 Tsudanuma, Narashino, Chiba 275-0016, Japan

## Abstract

Some atmospheric gases have been proposed as counter indicators to the presence of life on an exoplanet if remotely detectable at sufficient abundance (i.e., antibiosignatures), informing the search for biosignatures and potentially fingerprinting uninhabited habitats. However, the quantitative extent to which putative antibiosignatures could exist in the atmospheres of inhabited planets is not well understood. The most commonly referenced potential antibiosignature is CO, because it represents a source of free energy and reduced carbon that is readily exploited by life on Earth and is thus often assumed to accumulate only in the absence of life. Yet, biospheres actively produce CO through biomass burning, photooxidation processes, and release of gases that are photochemically converted into CO in the atmosphere. We demonstrate with a 1D ecosphere-atmosphere model that reducing biospheres can maintain CO levels of ∼100 ppmv even at low $H_2$ fluxes due to the impact of hybrid photosynthetic ecosystems. Additionally, we show that photochemistry around M dwarf stars is particularly favorable for the buildup of CO, with plausible concentrations for inhabited, oxygen-rich planets extending from hundreds of ppm to several percent. Since $CH_4$ buildup is also favored on these worlds, and because $O_2$ and $O_3$ are likely not detectable with the *James Webb Space Telescope*, the presence of high CO (>100 ppmv) may discriminate between oxygen-rich and reducing biospheres with near-future transmission observations. These results suggest that spectroscopic detection of CO can be compatible with the presence of life and that a comprehensive contextual assessment is required to validate the significance of potential antibiosignatures.

## 1. Introduction

There are currently over 3900 known exoplanets,[11] many of which are rocky and orbiting within the circumstellar habitable zone (HZ) of their host star (e.g., Kane et al. 2016). Transit and radial velocity (RV) studies have already uncovered rocky worlds in the HZ that would be amenable to future in-depth spectroscopic characterization, such as Proxima Centauri b (Anglada-Escudé et al. 2016), LHS 1140b (Dittmann et al. 2017), and one to three of the seven known planets orbiting TRAPPIST-1 (Gillon et al. 2017; Luger et al. 2017). The recently launched TESS telescope, together with ongoing ground-based transit and RV studies, will likely find many more potentially habitable exoplanets transiting nearby M- and K-type stars (Ricker et al. 2014; Sullivan et al. 2015). The next horizon in the search for life beyond the solar system will be the detailed characterization of exoplanet atmospheres, whether through transit spectroscopy of rocky planets with the *James Webb Space Telescope* (Deming et al. 2009; Stevenson et al. 2016; Krissansen-Totton et al. 2018a) or direct-imaging with future space and ground-based observatories (see the review in Fujii et al. 2018).

The science of defining and describing potential exoplanet biosignatures is an emerging but active field with several recent reviews available (Seager et al. 2012; Kaltenegger 2017; Kiang et al. 2018; Schwieterman et al. 2018b). Proposed strategies for life detection typically focus on techniques that would allow for a discrete affirmation of life beyond Earth, rather than on methodologies for positively identifying uninhabited environments, exploring the complexities surrounding ambiguous cases, or developing a statistical rather than case-by-case approach toward constraining the presence or absence of life. There are exceptions to this general rule, such as voluminous recent work exploring false positives for oxygen-based exoplanet biosignatures (abiotic processes that may be mistaken for life; see reviews in Harman & Domagal-Goldman 2018; Meadows 2017; Meadows et al. 2018b), hypothetical "false negatives" for life using Earth's history as an example (Reinhard et al. 2017; Olson et al. 2018b), and recently proposed statistical frameworks for the assessment of habitability and biosignatures (Bean et al. 2017; Catling et al. 2018; Walker et al. 2018).

---

[11] https://exoplanets.nasa.gov/

Antibiosignatures—or evidence indicating that life is *not* present—would in principle be useful for a variety of reasons. For example, if the overriding objective is to find evidence of life in an exoplanet spectrum, the remote detection of a convincing antibiosignature in a planet's atmosphere (or surface) would allow for the redirection of resources to more promising targets and would reduce the likelihood of false positive claims for life detection. Perhaps more impactful would be the positive identification of uninhabited planets, which could inform the prior probability of abiogenesis (origin of life) and provide an abiotic fiducial for examining the role of biology in planetary processes on Earth-like planets (see Cockell 2011, 2014a; Cockell et al. 2012, 2016). Antibiosignatures are also important to consider for solar system exploration (Neveu et al. 2018), though this is beyond our scope here.

Despite the possible advantages of developing robust antibiosignatures, the possible range and applicability of these signals are almost never comprehensively discussed in the literature. Even in cases where such complexities are addressed, the language employed can be confusing. For example, "false positives" for biosignatures are not necessarily antibiosignatures nor are they necessarily indicative of uninhabitable conditions. In other words, abiotic processes that produce putative biosignatures can be compatible with habitable environments. In addition, the presence of a confirmed antibiosignature need not necessarily imply that a planetary surface is uninhabitable—just that life is currently absent or limited in its energy consumption.

In this paper, we examine the common presumption that CO is a spectroscopic antibiosignature as a test case and find that spectroscopically detectable CO is compatible with an inhabited biosphere in some cases. We expand from this example to argue that there are no "smoking-gun" antibiosignatures, where a detection of a single gas above a prescribed threshold is definitive evidence for a sterile planet. Instead, considering environmental context and constraining the concentrations of putative antibiosignature gases are critical for proper interpretation.

In Section 2, we summarize what is known and what has been proposed about remote antibiosignatures with a focus on carbon monoxide (CO), since it is the most widely cited putative remotely detectable antibiosignature. In Section 3, we present a case study using a coupled 1D atmosphere-ecosphere model that demonstrates how CO can accumulate to levels exceeding 100 ppmv in an anaerobic biosphere that includes CO-consuming acetogens, potentially even at low volcanic $H_2$ fluxes providing that anoxygenic photosynthesizers can metabolize ferrous iron. In Section 4, we demonstrate, using a 1D photochemical model, that CO buildup is highly favored around late-type stars, particularly for Earth-like oxic planets with terrestrial biospheres and substantial surface CO sources. In these cases, CO mixing ratios may approach those of the equivalent abiotic case, showing that antibiosignature arguments drawn from the Earth–Sun case are not generally applicable. We summarize and conclude in Section 5.

## 2. Antibiosignatures: Evidence of Absence or Absence of Evidence?

An antibiosignature is often loosely defined as evidence of chemical free energy in the environment that is not currently being exploited by life. Catling et al. (2018) define an antibiosignature as "any substance, group of substances, or phenomenon that provides evidence against the presence of life." In a Bayesian probabilistic framework, Walker et al. (2018) define an "antibiosignature" as "an object, substance, and/or pattern that diminishes the likelihood the signal is generated by life, such that $P(\text{data}|\text{life})$ is less than in its absence (e.g., a given piece of contextual information $C$ is an antibiosignature if $P(\text{data}|\text{life}, C) < P(\text{data}|\text{life})$)." Here $P(\text{data}|\text{life})$ is the probability that the observed data are indicative of life, given the prior probability of life estimated from other contextual information, such as the position of the planet in the circumstellar habitable zone. In the above case, $C$ could be the detection of an antibiosignature gas.

Compared to the examination of potential planetary habitability and positive signs of life, remote antibiosignatures are deeply understudied and usually mentioned only as an aside in the context of a proposed biosignature or suite of biosignatures. One of the most commonly proposed general antibiosignatures is CO, because it builds up most favorably in the absence of water vapor (and a surface water ocean) and may be indicative of an unexploited source of free energy and reduced carbon (Ragsdale 2004; Zahnle et al. 2008; Gao et al. 2015; Wang et al. 2016; Catling et al. 2018). In our own solar system, the presence of CO has been proposed as an antibiosignature on Mars (Weiss et al. 2000) and calculations based on the observed CO abundance in the Martian atmosphere have been used to set limits on the potential size of a subsurface biosphere (Sholes et al. 2018). Indeed, the presence of CO in an exoplanet atmospheres has been recently proposed as both an antibiosignature (Wang et al. 2016) and, in its absence, as a key false positive discriminant for $CH_4$–$CO_2$ disequilibrium biosignatures in the anoxic atmospheres of terrestrial exoplanets similar to the Archean Eon (2.5–4.0 Ga) on Earth (Krissansen-Totton et al. 2018a, 2018b).

However, discussions of antibiosignatures have thus far failed to account for a number of key considerations such as macro or trace nutrient limitation and substantial differences in photochemistry for planets orbiting stars unlike the Sun. Without a quantitative exploration of these factors—or even a recognition of possible limitations in the absence of quantitative analyses—antibiosignature concepts will be limited in their applicability. However, some existing studies provide a starting point for understanding quantitative limits on antibiosignature gases, particularly CO. For planets with biospheres, we draw particular attention to the work of Kharecha et al. (2005) who used a coupled ecosphere-atmosphere (photochemical) model to explore the plausible atmospheric chemistry of the early Archean Earth before the advent of oxygenic photosynthesis (>3.0 Ga). They considered suites of metabolisms including methanogenesis (net reaction $CO_2 + 4H_2 \rightarrow CH_4 + 2H_2O$), acetogenesis (net reaction $4CO + 2H_2O \rightarrow 2CO_2 + CH_3COOH$), and $H_2$-based photosynthesis (net reaction $CO_2 + 2H_2 + h\nu \rightarrow CH_2O + H_2O$).

Broadly, Kharecha et al. (2005) found that methanogenesis-based ecosystems without CO-consuming acetogens may allow atmospheric CO to build up to high levels (>10% by volume). This is partly because biogenic $CH_4$ is photochemically oxidized to CO, which itself is dependent on the $H_2$ flux through the activity of methanogens that convert volcanic $H_2$ to biogenic $CH_4$. In this oxygen-free case, the major sink of CO is hydroxyl radicals (OH) liberated by the photolysis of water ($H_2O + h\nu \rightarrow H + OH$). With only this sink for CO, the gas can build to concentrations comprising a significant fraction of the atmosphere, and a CO "runaway" can occur at sufficiently high $H_2$ fluxes or low $H_2O$ concentrations (see also Gao et al. 2015). However, Kharecha et al. (2005) argued that on Earth the metabolic pathway for CO consumption is likely as ancient as methanogenesis, so it should also be considered in determining plausible CO concentrations. In this case (i.e., on the Archean Earth), the major sink for CO is deposition into the ocean and consumption by acetogens and CO can potentially be drawn down to much lower atmospheric abundances.

The maximum rate of CO deposition can be quantified by assuming that the surface ocean has a CO concentration of zero, equivalent to assuming that acetogens immediately consume any dissolved CO. In this case, CO is only limited by the rate at which it can transfer from the atmosphere into the ocean, which can be described quantitatively through a deposition velocity. Steady-state CO levels under these conditions vary depending on the assumed $H_2$ flux, but can exceed 100 ppmv ($\sim 1 \times 10^{-4}$ v/v) at high $H_2$ fractions ($fH_2 \sim 5000$ ppmv). Though this atmospheric abundance is well below the estimated "runaway" levels discussed above, it is approximately three orders of magnitude higher than the modern CO level of $\sim 0.1$ ppmv (100 ppb; $1 \times 10^{-7}$ v/v). This relationship suggests that a more accurate understanding of the limits of CO buildup for the Earth–Sun case is essential for properly bounding scenarios where habitable planets orbit different star types, which can result in different photochemical lifetimes and steady-state concentrations for CO despite similar surface fluxes.

In the next section, we examine atmospheric CO abundance in the presence of primitive anoxic biospheres with $H_2$-based and coupled $H_2$- and $Fe^{2+}$-based anoxygenic photosynthetic biospheres. We find maximum CO concentrations similar to those reported in Kharecha et al. (2005) and a similar scaling between volcanic $H_2$ fluxes and steady-state atmospheric CO abundance. However, we also note that the hybrid photosynthetic ($H_2$ + $Fe^{2+}$) cases can potentially generate relatively high CO at much lower $H_2$ fluxes than would be inferred from Kharecha et al. (2005) when fluxes of $Fe^{2+}$ to the ocean are high. We also find that steady-state atmospheric CO abundance scales directly with overall biospheric productivity, such as that for primitive photosynthetic biospheres that are highly productive "intermediate" atmospheric CO abundance may actually serve as a positive biosignature.

## 3. Example 1: Biogenic CO Accumulation on an Anaerobic Archean Earth

Here we present self-consistent CO and $CH_4$ mixing ratios from a coupled photochemistry ecosphere model originally developed for examining the impact of anoxygenic photosynthesis on the climate and atmosphere of the early Archean Earth (Ozaki et al. 2018). The photochemical component of the model includes 73 chemical species and 359 reactions and is appropriate for an anoxic atmosphere with no ground-level $O_2$ flux from oxygenic photosynthesis. The ecosphere component includes two cases. In Case 1, the metabolisms included are $H_2$ consuming anoxygenic photosynthesis, CO-consuming acetogenesis, organic matter fermentation, and acetotrophic methanogenesis. In Case 2, a "hybrid" photosynthetic biosphere is constructed by adding $Fe^{2+}$-consuming anoxygenic photosynthesis and dissimilatory Fe(III)-reducing bacteria to the four metabolisms in Case 1 (Figure 1). We note that model results are generally similar for alternative configurations that employ alternative pathways of $CH_4$ production that do not involve acetoclastic methanogenesis (Ozaki et al. 2018). Full model details and boundary conditions for individual model runs presented here can be found in Ozaki et al. (2018).

Figure 1 shows atmospheric $pCH_4$ and $pCO$ as functions of the volcanic $H_2$ outgassing flux. (For reference, a flux of $10^{12}$ moles of $H_2$ per year is equivalent to an instantaneous surface flux of $\sim 3.8 \times 10^9$ molecules cm$^{-2}$ s$^{-1}$). In Case 1, both $CH_4$ and CO abundances are strong functions of the volcanic outgassing rate. In this case, higher $H_2$ fluxes allow higher productivity of anoxygenic phototrophs, which increases the rate of methanogenesis and subsequently the CO derived from the photochemical processing of $CH_4$ into CO in the atmosphere. Since CO is also derived from the photolysis of $CO_2$ ($CO_2 + h\nu \rightarrow CO + O$), $pCO$ is also dependent on the $CO_2$ mixing ratio. This dependence is more impactful at low $H_2$ fluxes, where $CO_2$ photolysis is a proportionally greater source of CO relative to $CH_4$ photolysis. (Importantly, this simple model neglects direct sources of CO from volcanism, which are small, and biomass burning, since that source is inconsistent with an anoxic, primarily marine biosphere.) In Case 2, photoferrotrophy is also included in the model, leading to strong, nonlinear amplification of atmospheric $CH_4$ abundance even at relatively low volcanic $H_2$ fluxes (Figure 1(b)). Photoferrotrophy fixes $CO_2$ into organic matter by using light energy and reduced iron ($Fe^{2+}$) as an electron donor. Since the reducing Archean oceans were replete with $Fe^{2+}$, photoferrotrophy likely played an extensive role in the productivity of ancient oceans (Camacho et al. 2017). Ozaki et al. (2018) suggested that metabolic diversity in the Archean could partly explain the warm early Archean climate state suggested by the geologic record, but an important corollary to this observation that was not noted by Kharecha et al. (2005) or Ozaki et al. (2018) is that this metabolic diversity may also facilitate relatively high levels of CO at comparatively low volcanic $H_2$ fluxes.

It is important to emphasize that the capacity of our primitive photosynthetic biosphere to sustain elevated atmospheric CO abundances at low volcanic $H_2$ flux scales with the flux of $Fe^{2+}$ to surface environments. For example, the benchmark models shown in Figure 1(b) assume an $Fe^{2+}$ flux of 80 Tmol Fe yr$^{-1}$. Ozaki et al. (2018) estimated an upper bound on the $Fe^{2+}$ flux from high-temperature ($\sim 350\,°C$), on-axis hydrothermal systems on the Archean Earth of $\sim 10$–15 Tmol Fe yr$^{-1}$, to which can be added up to $\sim 10$ Tmol Fe yr$^{-1}$ (Raiswell 2006) assuming that all reactive Fe currently weathered from the upper crust would have been delivered as a dissolved species in a strongly reducing ocean-atmosphere system. Additional potential fluxes of $Fe^{2+}$ from low-temperature, ridge flank hydrothermal systems in an anoxic ocean are unconstrained, but could be large. Potential limits on the flux of reduced Fe from the solid planet for exoplanet scenarios are also not well constrained. The atmospheric abundance of CO at a given volcanic $H_2$ flux will also respond to background $pCO_2$ and the burial efficiency of organic matter in marine sediments.

In any case, it seems clear that for reasonable planetary fluxes of $H_2$ and $Fe^{2+}$ our primitive biosphere is capable of maintaining atmospheric CO abundances of order $10^{-4}$ bar. Interestingly, atmospheric CO abundance in our model scales directly with overall biospheric productivity, with the somewhat counterintuitive result that very low CO abundance in the atmosphere only results when overall biospheric productivity is extremely low. Nevertheless, these results are applicable for an early Archean Earth-like planet orbiting a Sun-like star and for cases where the direct source of CO into the atmosphere (biotic or abiotic) is relatively small. In the next section, we test the general relationships of $pCO$ and $pCH_4$ to $CH_4$ and CO fluxes for both modern Earth-like (oxic) and Archean-like (anoxic and reducing) cases and for planets orbiting both a Sun-like star and a late M dwarf like Proxima Centauri. We also test a smaller set of $pCO$ relationships to CO flux for a larger set of stellar hosts (FGKM).

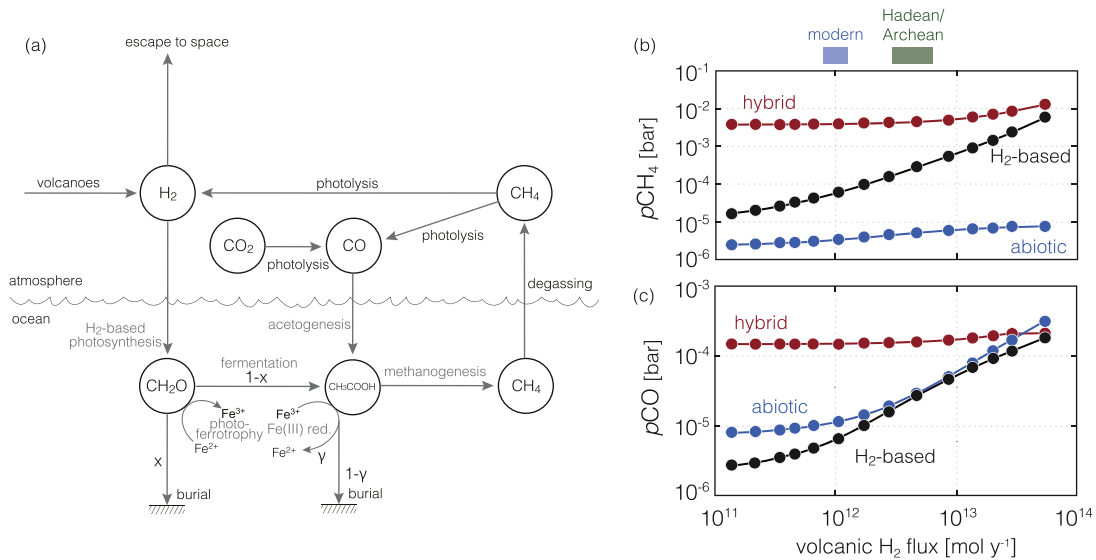

**Figure 1.** Atmospheric methane ($CH_4$) and carbon monoxide (CO) in our coupled ecosphere-atmosphere model. Shown on the left is (a) a schematic depiction of the model structure. Shown on the right are (b) atmospheric $CH_4$ and (c) CO as functions of volcanic $H_2$ flux for abiotic (blue), $H_2$-based (black), and "hybrid" (red) biospheres, the latter of which includes both $H_2$-based photosynthesis and Fe-based photosynthesis. Also shown in (b) and (c) are modern and estimated Hadean/Archean volcanic $H_2$ fluxes (top). Results shown in (b) and (c) assume $pCO_2 = 0.1$ bar, while hybrid calculations assume an $Fe^{2+}$ flux of 80 Tmol yr$^{-1}$.

## 4. Example 2: Photochemical CO Buildup on Inhabited Planets Orbiting Late-type Stars

In this section, we test the sensitivity of $pCO$ to CO fluxes for hypothetical oxic and anoxic/reducing Earth-like planets orbiting at the inner edge of the habitable zone of the Sun and Proxima Centauri (a late M star). Note that to control for all other parameters (e.g., planet size, position within the habitable zone, and surface gravity) this hypothetical planet is not meant to exactly represent the recently discovered planet Proxima Centauri b (Anglada-Escudé et al. 2016). We additionally test $pCH_4$ relative to $CH_4$ fluxes for the same scenarios, because high $CH_4$ concentrations have been proposed as a biosignature in the presence of $CO_2$ and absence of significant CO (Krissansen-Totton et al. 2018a, 2018b). Our goal here is not to completely explore all photochemical scenarios for CO buildup on planets around other stars, but rather we illustrate a small set of cases where high CO may exist in the atmosphere of an inhabited planet in the presence of high concentrations of (biologically sourced) $CH_4$. We also generally test the capacity for CO build-up for Earth twins orbiting a range of (FGKM) stellar hosts, including Sigma Boötis, Epsilon Eridani, AD Leo, GJ 876, and TRAPPIST-1.

We use the 1D photochemical code contained within the publicly available *Atmos*[12] model. *Atmos* is derived from the photochemical code originally developed by the Kasting group (e.g., Kasting et al. 1979; Pavlov et al. 2001) but with several additions and modifications. Recently, the code was used to calculate photochemically self-consistent atmospheres for the Archean Earth and hazy planets orbiting other stars (Arney et al. 2016, 2017) and for calculating self-consistent trace gas abundances in the atmosphere of Proxima Centauri b (Meadows et al. 2018a). Our stellar spectra are sourced from the Virtual Planetary Laboratory and have all been used in prior publications (Segura et al. 2003, 2005; Lincowski et al. 2018; Meadows et al. 2018a).

Our procedures are as follows. For our oxic test cases, we begin by using a modern Earth template in a previously converged state (i.e., reproducing modern composition, 78% $N_2$, 21% $O_2$ v/v, 360 ppm $CO_2$) and use the model to calculate the net CO and $CH_4$ fluxes required to sustain the observed volume mixing ratios for those gases in the modern atmosphere. We then use these fluxes as a bottom boundary condition to calculate self-consistent trace gas concentrations throughout a 1D atmospheric column, while altering the input stellar spectrum or adjusting our flux boundaries by an integer multiple of its original value (Procedure 1). This same approach has been used by a range of authors in previous studies (Segura et al. 2005; Rugheimer et al. 2013, 2015; Meadows et al. 2018a). We use an additional procedure (Procedure 2) to reflect the most conservative conditions for CO buildup in the atmosphere, which assumes both a CO flux distributed over the bottom kilometer of the atmosphere and a maximum surface deposition boundary condition of CO of $1.2 \times 10^{-4}$ cm s$^{-1}$, which is the maximum biotic rate assuming that CO dissolved in the ocean is immediately consumed (Kharecha et al. 2005). (Importantly, and for reasons discussed below, both Procedures 1 and 2 produce identical results for the modern, i.e., oxygen-rich, Earth–Sun configuration).

We alter our CO and $CH_4$ fluxes independently, while fixing the unaltered gas flux at the modern value. For our "anoxic" scenarios we adopt boundary conditions consistent with an Archean Earth with 2% $CO_2$ and thin or no haze

---

[12] https://github.com/VirtualPlanetaryLaboratory/atmos

Table 1
Oxic Boundary Conditions

| Chemical Species[a] | Deposition Velocity (cm s$^{-1}$) | Flux (molecules cm$^{-2}$ s$^{-1}$) | Mixing Ratio |
|---|---|---|---|
| O | 1 | ... | ... |
| $O_2$ | ... | ... | 0.21 |
| $N_2$ | ... | ... | 0.78 |
| $CO_2$ | ... | ... | $3.6 \times 10^{-4}$ |
| $H_2O$ | ... | ... | fixed[b] |
| H | 1 | ... | ... |
| OH | 1 | ... | ... |
| $HO_2$ | 1 | ... | ... |
| $H_2$ | $2.4 \times 10^{-4}$ | ... | $5.3 \times 10^{-7}$ |
| CO | $0 - 1.2 \times 10^{-4}$ | variable | ... |
| HCO | 1 | ... | ... |
| $H_2CO$ | 0.2 | ... | ... |
| $CH_4$ | 0 | variable | ... |
| $CH_3$ | 1 | ... | ... |
| NO | $3 \times 10^{-4}$ | $1 \times 10^9$ | ... |
| $NO_2$ | $3 \times 10^{-3}$ | ... | ... |
| HNO | 1 | ... | ... |
| $H_2S$ | ... | $1 \times 10^8$ | ... |
| $SO_2$ | 1 | $1 \times 10^9$ | ... |
| $H_2SO_4$ | 1 | ... | ... |
| HSO | ... | ... | ... |
| $O_3$ | 0.07 | ... | ... |
| $HNO_3$ | 0.2 | ... | ... |
| $N_2O$ | ... | ... | $3.1 \times 10^{-7}$ |
| $HO_2NO_2$ | 0.2 | ... | ... |

**Notes.**
[a] Species included in the photochemical scheme with a deposition velocity and flux of 0 include: $C_2H_6$, HS, S, SO, $S_2$, $S_4$, $S_8$, $SO_3$, OCS, $S_3$, N, $NO_3$, and $N_2O_5$.
[b] The $H_2O$ profile is fixed to an Earth average.

Table 2
Anoxic Boundary Conditions

| Chemical Species[a] | Deposition Velocity (cm s$^{-1}$) | Flux (molecules cm$^{-2}$ s$^{-1}$) | Mixing Ratio |
|---|---|---|---|
| O | 1 | ... | ... |
| $O_2$ | $1.4 \times 10^{-4}$ | ... | ... |
| $N_2$ | ... | ... | 0.80 |
| $CO_2$ | ... | ... | 0.02 |
| $H_2O$ | ... | ... | fixed[b] |
| H | 1 | ... | ... |
| OH | 1 | ... | ... |
| $HO_2$ | 1 | ... | ... |
| $H_2$ | $2.4 \times 10^{-4}$ | $1 \times 10^{10}$ | ... |
| CO | $1.2 \times 10^{-4}$ | variable | ... |
| HCO | 1 | ... | ... |
| $H_2CO$ | 0.2 | ... | ... |
| $CH_4$ | 0 | variable | ... |
| $CH_3$ | 1 | ... | ... |
| NO | $3 \times 10^{-4}$ | $1 \times 10^9$ | ... |
| $NO_2$ | $3 \times 10^{-3}$ | ... | ... |
| HNO | 1 | ... | ... |
| $O_3$ | 0.07 | ... | ... |
| $HNO_3$ | 0.2 | ... | ... |
| $H_2S$ | 0.02 | $3.5 \times 10^8$ | ... |
| HSO | 1 | ... | ... |
| $H_2SO_4$ | 1 | ... | ... |
| $SO_2$ | 1 | $3.5 \times 10^9$ | ... |

**Notes.**
[a] Species included in the photochemical scheme with a deposition velocity and flux of 0 include: N, $C_3H_2$, $C_2H_6$, $C_3H_3$ $CH_3C_2H$, $CH_2CCH_2$, $C_3H_5$, $C_2H_5CHO$, $C_3H_6$, $C_3H_7$, $C_3H_8$, $C_2H_4OH$, $C_2H_2OH$, $C_2H_5$, $C_2H_4$, CH $CH_3O_2$, $CH_3O$, $CH_2CO$, $CH_3CO$, $CH_3CHO$, $C_2H_2$, $(CH_2)_3$, $C_2H$, $C_2$, $C_2H_3$, HCS, $CS_2$, CS, OCS, S, HS, $SO_3$, $S_2$, and SO.
[b] The $H_2O$ profile is fixed to an Earth average.

(Arney et al. 2016, 2017). For both CO and $CH_4$ we step through eight orders of magnitude in surface gas flux, from $\sim 10^6$ to $\sim 10^{14}$ molecules cm$^{-2}$ s$^{-1}$. The temperature-pressure and water vapor mixing ratio profiles for both oxic and anoxic cases are consistent with a planet with a surface temperature of 288 K. Our complete list of boundary conditions for our oxic cases are given in Table 1, and our anoxic boundary conditions are given in Table 2.

The procedures described above produce accurate results when applied to the modern Earth, with some minor exceptions and caveats. For example, the photochemical model slightly overestimates fluxes for $CH_4$ and CO relative to literature estimates, which was also noted by Segura et al. (2005). For $CH_4$, we derive a flux of $1.5 \times 10^{11}$ molecules cm$^{-2}$ s$^{-1}$ using the photochemical model, which can be converted to $\sim 640$ Tg yr$^{-1}$. This result compares favorably to a modern estimated global methane flux of 500–600 Tg yr$^{-1}$ (Dlugokencky et al. 2011). For CO, we derive a molecular flux $3.0 \times 10^{11}$ molecules cm$^{-2}$ s$^{-1}$, which is equivalent to $\sim 1280$ Tg yr$^{-1}$. This CO flux estimate compares relatively well to a recent empirical estimate of 888 Tg yr$^{-1}$ with a confidence interval of 745.67–1112.80 Tg yr$^{-1}$ (Zhong et al. 2017). These slight overestimations are likely due to the limited accuracy of the one-dimensional model, which cannot account for spatial heterogeneity (e.g., ocean versus land) or seasonal changes. We also use the current mixing ratios of these two gases (to match publicly available photochemical templates and facilitate reproduction of results), which are influenced by anthropogenic emissions. However, the current concentration of $CH_4$ and CO in the atmosphere are not unprecedented in recent (Phanerozoic) Earth history, although they may be significantly higher than immediate preindustrial concentrations (Olson et al. 2018b).

It is important to emphasize that the lower boundary conditions for CO are significantly different for the oxygen-rich modern Earth compared to the idealized oxygen-poor Archean-like scenarios explored in Section 2. For example, today the modern ocean is a net source of CO from photooxidation of organic matter and production by phytoplankton rather than a net sink for CO—despite the presence of CO-consuming acetogens (Conrad et al. 1982; Blomquist et al. 2012; Conte et al. 2018). In our photochemical tests, the necessary CO flux required to reproduce the empirical mixing ratio is essentially insensitive to the assumed ocean deposition velocity because terrestrial (land-based) CO production is so much greater than the maximum deposition rate and because of short atmospheric CO lifetimes in the atmosphere. However, for planets orbiting late-type stars the atmospheric lifetimes of CO are much greater, thus increasing the CO partial pressures and the impact of high deposition fluxes. Consequently, these two procedures yield significantly different estimates. In this case, given an assumed molecular CO flux the steady-state concentration should lie within the ranges bracketed by our Procedure 1 and Procedure 2 assumptions. More accurate calculations will require the development of more sophisticated and generalized 3D photochemical models.

For both the oxic and anoxic planets, higher concentrations of CO and $CH_4$ are predicted for the hypothetical planet orbiting Proxima Centauri compared to estimates for the Sun (Figures 2 and 3 and ). These higher abundance estimates are primarily due to the lower atmospheric OH concentrations in planetary atmospheres around Proxima Centauri. The hydroxyl radical (OH) is the primary sink for both CO and CH$_4$ through the reactions CO + OH → CO$_2$ + H and CH$_4$ + OH → CH$_3$ + H$_2$O. For oxic atmospheres, OH is primarily sourced through the reaction H$_2$O + O($^1$D) → 2OH. The singlet oxygen is derived from tropospheric ozone through the reaction O$_3$ + $h\nu$ ($\lambda <$ 320 nm) → O($^1$D) + O$_2$. Thus, the sequence of reactions that ultimately result in tropospheric OH is mediated by NUV radiation that is substantially less plentiful from M-stars such as Proxima Centauri because of lower photosphere blackbody temperatures, with the result that OH production is much less favored and consequently the sinks for CO and CH$_4$ are much less efficient (Segura et al. 2005). In anoxic atmospheres, OH is derived primarily from photolysis of H$_2$O (H$_2$O + $h\nu$ [$\lambda <$ 200 nm] → OH + H), which is partially shielded by overlying CO$_2$. The FUV radiation of Proxima Centauri is highly concentrated near the Ly$\alpha$ wavelengths ($\sim$121.6 nm), which makes this self-shielding more effective and reduces the OH abundance for the Proxima anoxic atmospheres versus the solar anoxic atmospheres.

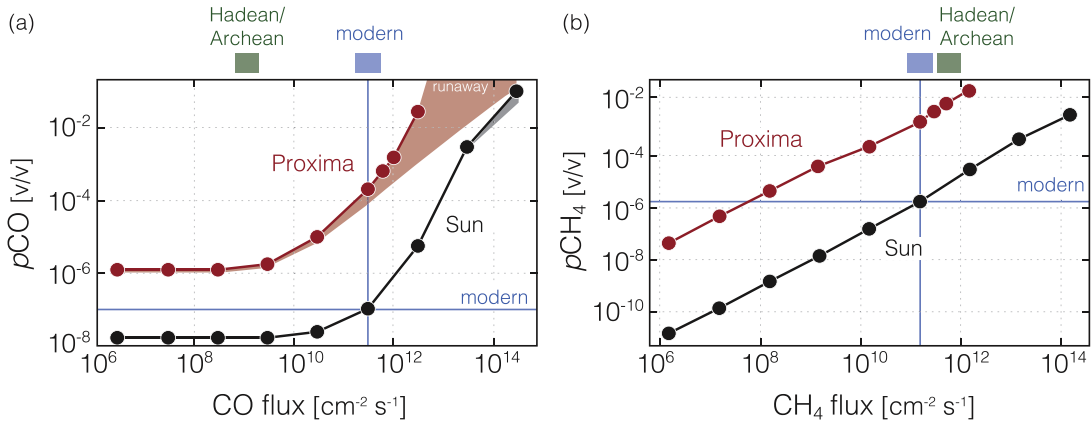

**Figure 2.** Atmospheric mixing ratios of CO (a) and CH$_4$ (b) as a function of surface molecular fluxes for oxygen-rich, modern Earth-like atmospheres. Results are shown for both the Sun (black) and the M-star Proxima Centauri (red). Horizontal line shows mixing ratios for each gas in the modern atmosphere, while modern and estimated Hadean/Archean fluxes of CO and CH$_4$ are shown above each panel. The shaded ranges in (a) depict results for models assuming only an upward molecular flux of CO (upper end) and assuming both an upward molecular flux and the maximum possible deposition velocity of $1.2 \times 10^{-4}$ cm s$^{-1}$ (lower end). Note that for the highest surface fluxes around Proxima Centauri the atmosphere enters a "CO runaway." Atmospheric $p$CO$_2$ is fixed at 360 ppm ($\sim 3.6 \times 10^{-4}$ bar) in all calculations.

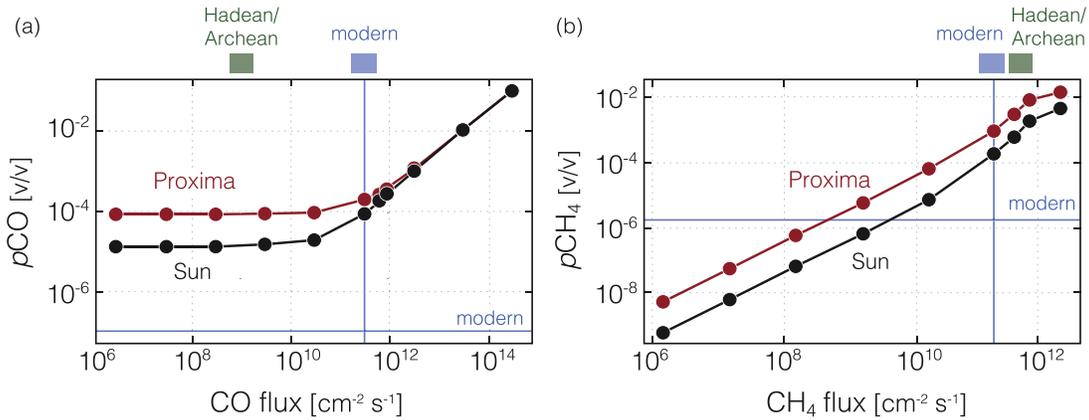

**Figure 3.** Atmospheric mixing ratios of CO (a) and CH$_4$ (b) as a function of surface molecular fluxes for oxygen-poor, Archean Earth-like atmospheres. Results are shown for both the Sun (black) and the M-star Proxima Centauri (red). Horizontal line shows mixing ratios for each gas in the modern atmosphere, while modern and estimated Hadean/Archean fluxes of CO and CH$_4$ are shown above each panel. Atmospheric $p$CO$_2$ is fixed at 0.02 bar in all calculations.

Figure 2 shows resulting CO and CH$_4$ mixing ratios for our oxic (modern Earth-like) planet. Note that at low CO fluxes, the atmospheric CO is predominately sourced from CO$_2$ photolysis, with a weak relationship to CO$_2$ mixing ratio due to photochemical self-shielding (not shown). At these low CO fluxes the planet orbiting the Sun has a CO mixing ratio of $<2 \times 10^{-8}$, while the planet orbiting Proxima Centauri has a CO mixing ratio two orders of magnitude higher ($>$1 ppm). Notably, at the same CO flux (3 $\times 10^{11}$ molecules cm$^{-2}$ s$^{-1}$) as derived from the modern Earth, the steady-state $p$CO concentration for a planet orbiting at the inner habitable zone boundary of Proxima Centauri is over three orders of magnitude higher ($>$200 ppm for Procedure 1 versus $\sim$0.1 ppmv for the Sun case). At a CO flux of three times modern the CO mixing ratio would be $>$1500 ppm, and for 10 times the modern flux $p$CO is approximately 2.7% (Procedure 1). The latter matches or exceeds the maximum predicted CO concentration on a habitable but uninhabited planet with an active hydrological cycle and limited NO$_x$ production from lightning (Harman et al. 2015, 2018), even though our model assumes both to be in operation. For the more conservative lower boundary conditions of Procedure 2, the predicted CO concentrations are 59 ppmv, 200 ppmv, and 780 ppmv, for surface CO fluxes of 1, 3, and 10 times modern, respectively. Atmospheric $CH_4$ concentrations in our oxic case are also a strong function of the molecular flux and highly sensitive to the stellar spectrum. At $CH_4$ fluxes characteristic of the modern Earth the $CH_4$ concentration on the Earth twin orbiting Proxima Centauri would be 1890 ppm (compared to 1.7 ppm in the modern atmosphere), with significant spectroscopic and climatic implications (see, e.g., Meadows et al. 2018a).

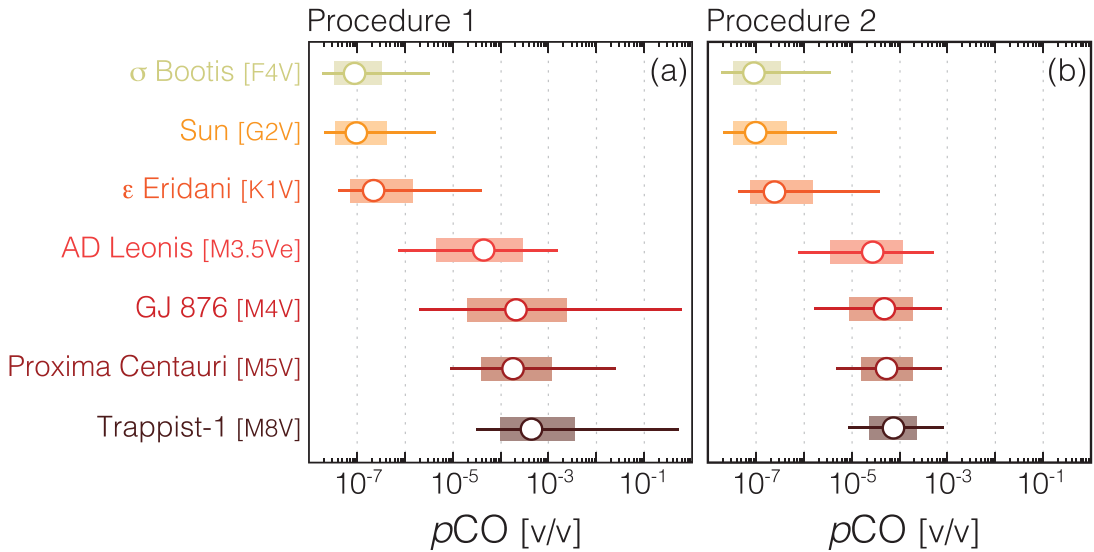

**Figure 4.** Steady-state ground-level atmospheric mixing ratios of CO for modern Earth-like (78% $N_2$, 21% $O_2$, 360 ppmv $CO_2$) planets around a range of stellar hosts. Open circles show results for the modern surface CO flux, while ranges show results of increasing/decreasing this flux by a factor of 3 (shaded bars) or 10 (horizontal lines). Procedure 1 (a) shows results from models assuming only a net upward molecular flux of CO, while Procedure 2 (b) shows results from models assuming both an upward molecular flux and the maximum possible deposition velocity of $1.2 \times 10^{-4}$ cm s$^{-1}$. Note the log scale.

Figure 3 shows the relationships between CO and $CH_4$ fluxes and atmospheric abundances for the anoxic Archean Earth scenarios. For CO, the primary difference between oxic and anoxic cases is the baseline-level CO mixing ratio from only $CO_2$ photolysis. The CO concentration at low CO fluxes is substantially higher for the anoxic planets than for the oxic ones (~10 ppm for the Sun, and ~100 ppm for Proxima Centauri). This difference is due to the higher abundance of reducing gases such as $H_2$, which promotes CO production in the atmosphere. At these low CO fluxes, the differences in CO concentrations between the solar case and the Proxima case are driven by different efficiencies of CO production through photolysis. At high CO fluxes this difference disappears because the CO photochemical lifetimes are similar in these anoxic atmospheres. However, near-modern fluxes of CO are likely implausible in anoxic atmospheres because biomass burning, a significant source of CO on the modern Earth, would be chemically impossible without free oxygen. Therefore, somewhat counterintuitively, simultaneously high abundances of $CH_4$ (>0.1%) and CO (>>100 ppm) may be more likely in oxic atmospheres than anoxic ones for planets orbiting M dwarf stars.

Figure 4 shows estimated CO concentrations predicted in modern Earth-twin atmospheres (78% $N_2$, 21% $O_2$, 360 ppm $CO_2$) for a range of FGKM stars, but over a more limited range of CO fluxes from one-tenth to 10 times the modern net flux. These results suggest that CO concentrations will be low (<1–10 ppmv) for most FGK stars, providing they do not have extremely high surface CO emission. However, we predict that planets orbiting in the habitable zones of M dwarf stars could have significant CO concentrations (>100 ppmv) for near-modern net CO fluxes with a general trend toward higher CO concentrations with decreasing stellar effective temperatures. There is essentially no difference in calculated CO concentrations for FGK stars between Procedure 1 and Procedure 2; however, this difference is significant for the M dwarf cases. For example, we predict that the CO concentration of an Earth-twin orbiting TRAPPIST-1 (with the modern surface CO flux of $3 \times 10^{11}$ molecules cm$^{-2}$ s$^{-1}$) would be ~500 ppmv with Procedure 1, but only ~80 ppmv with Procedure 2. In both cases the values are substantially greater than the present Earth concentration of 0.1 ppmv.

It is important to consider the impact of the assumptions used to calculate the mixing ratio profile of water vapor (80% saturation and a 288 K surface temperature). Colder temperatures would result in lower water vapor mixing ratios, and therefore even lower OH concentrations. This reduction in the primary photochemical CO sink could increase CO concentrations even further than the values given here. Colder temperatures, and therefore higher CO/$CH_4$ concentrations, could result from larger distances from the host star and therefore lower instellations (Grenfell et al. 2007) or from lower abundances of greenhouse gases such as $CO_2$. Lower $H_2O$ mixing ratios may also be found on mostly dry planets, perhaps interior to the traditional habitable zone (Abe et al. 2011; Zsom et al. 2013). Alternatively, the consumption of CO in the ocean may not rise to the biological limit (Case 2) if acetogens are limited by major or trace nutrients. Future work will be required to explore these possibilities.

To contextualize the results shown in Figures 2–4 in terms of detectability, we use the line-by-line radiative transfer model SMART (Meadows & Crisp 1996; Crisp 1997; Robinson et al. 2011; Robinson 2017) to simulate the transmission spectrum of a photochemically self-consistent Earth-twin orbiting an M dwarf star (Figure 5). We show these results for the 1–5 $\mu$m spectral region, which can be characterized with James Webb Space Telescope's (JWST) NIRISS and/or NIRSPEC instruments (Greene et al. 2016). We use the same series of flux conditions for an Earth-twin orbiting Proxima Centauri ($R = 0.141\, R_\odot$) as shown in Figure 4: 0, 0.1, 0.3, 1, 3, and 10 times the modern CO surface flux of $3 \times 10^{11}$ molecules cm$^{-2}$ s$^{-1}$. We also assume a surface $CH_4$ flux of $1.5 \times 10^{11}$ molecules cm$^{-2}$ s$^{-1}$ for all cases. We show these results for the permissive Procedure 1 boundary conditions (Figure 5(a)) and the more conservative Procedure 2 boundary conditions (Figure 5(b)). We also compare these results to an Archean-Earth atmosphere (Figure 5(c)). Our resulting spectra show that the 4.6 μm CO band is highly sensitive to surface CO flux and can match the transit depths of other significant bands such as $CH_4$ and CO for CO fluxes of 1–10 times modern. Intriguingly, for both the oxic (modern Earth) and anoxic (Archean Earth) cases, $CH_4$ bands are dominant throughout the spectrum and have a similar magnitude, suggesting that the determinations of the atmospheric oxidation state of terrestrial planets orbiting M dwarfs may be difficult with transmission spectra alone.

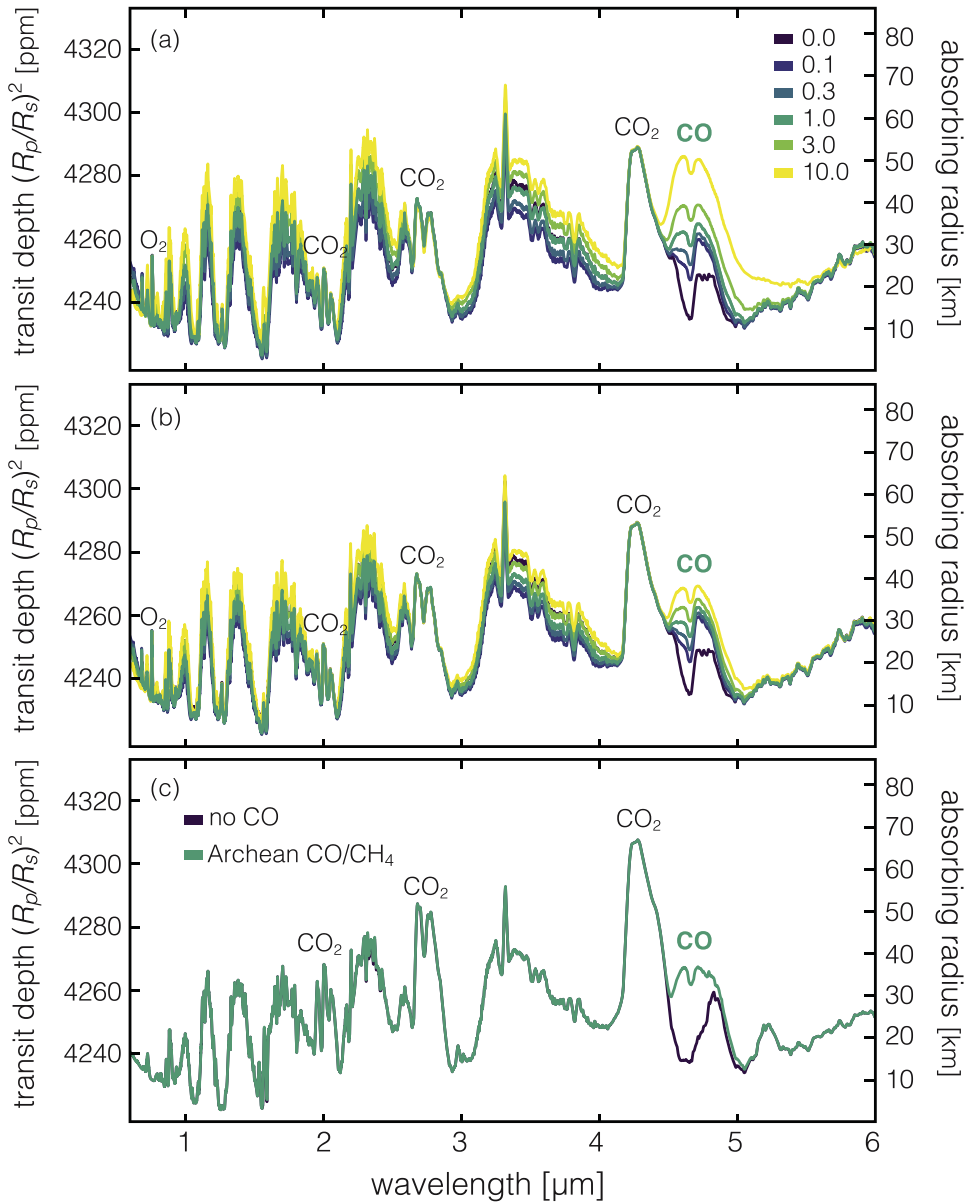

**Figure 5.** Simulated transmission spectra for major (CO, $CH_4$, $CO_2$) features in oxygen-rich, modern Earth-like atmospheres in the HZ of an M5V star (a), (b) and an oxygen-poor, Archean Earth-like atmosphere (c). Colors in (a) and (b) correspond to the magnitude of the assumed surface molecular CO flux scaled to that of the modern Earth (e.g., modern = 1.0). Results in (a) are from models assuming only an upward molecular flux of CO, while those in (b) are from models assuming both an upward molecular flux and the maximum possible ocean deposition velocity of $1.2 \times 10^{-4}$ cm s$^{-1}$. Unlabeled features are due to $CH_4$. Results for (c) assume mixing ratios of 885 ppm for $CH_4$, 116 ppm for CO, and 1% $CO_2$, levels compatible with the proxy record and photochemical modeling for Archean Earth (Arney et al. 2016; Olson et al. 2018b). Transit depth is calculated based on a star with R = 0.141 $R_\odot$ (i.e., an M5V star identical in radius to Proxima Centauri). $CH_4$ bands differ slightly in magnitude between CO flux scenarios in panels (a) and (b) because some CO is photochemically processed into $CH_4$ in the atmosphere.

## 5. Discussion and Conclusions

Previous work has suggested that detectable CO would be an antibiosignature, because it would be expected to be drawn down to low levels in the presence of life. Furthermore, the absence of CO could be a good check on $CH_4$/$CO_2$ disequilibrium biosignatures by helping to rule out a geologic source of $CH_4$ (Krissansen-Totton et al. 2018a). However, our work shows that the combination of biological sources of CO, molecular flux limitations imposed by the ocean-atmosphere interface, and photochemistry around late-type stars may produce simultaneously high levels of $CO_2$, $CH_4$, and CO even for planets with a productive biosphere. This possibility is most relevant for oxygen-rich planets like the modern Earth, which could produce high CO fluxes through biomass burning and photooxidation of dissolved organic matter in the surface ocean. Determining the oxidation states of such atmospheres may be difficult in the near-term, because it is likely *JWST* will be unable to detect $O_2$ or $O_3$ even with 10+ transits (Krissansen-Totton et al. 2018a). Thus, such planets may appear to be Archean-like Earths given the high $CH_4$ values possible for oxic planets orbiting M dwarfs (Segura et al. 2005; Rugheimer et al. 2015; Meadows et al. 2018a) with no way to distinguish between anoxic and oxic atmospheres without complementary observations by future ground- or space-based telescopes that can constrain $O_2$ or $O_3$ levels.

Our results introduce some important caveats to previous suggestions for interpreting CO in planetary atmospheres, particularly for planets orbiting M dwarfs. For example, our demonstration that high CO may be achieved on inhabited planets means that while simultaneously high $CO_2$ and $CH_4$ with little CO is still a compelling biosignature, ambiguous scenarios with high levels of all three gases also exist, and these are mostly relevant for transit transmission observations of habitable zone planets orbiting M dwarf stars. Arguments regarding threshold $CH_4$ levels that are incompatible with abiotic $CH_4$ outgassing rates are in principle still valid (Krissansen-Totton et al. 2018b), but the proposal that CO provides a check on abiotic versus biological origins of $CH_4$ is weakened by our results given likely near-future capabilities. Meanwhile, the photochemical "false positive" scenarios that lead to abiotic buildup of $O_2/O_3$ and CO are incompatible with large $CH_4$ mixing ratios (Domagal-Goldman et al. 2014) and high CO oxygen false positives are less favorable for FGK stars. The primary targets for future direct-imaging biosignature surveys (Bolcar et al. 2017; Martin et al. 2018; Roberge & Moustakas 2018) are thus not expected to have significant levels of $O_2$ and $CH_4$ and low levels of CO without the presence of life.

We stress that atmospheres with high CO present in a stochiometric ratio with O consistent with $CO_2$ photolysis may indeed place limits on the size of a biological CO sink, thus serving as an antibiosignature as previously suggested (Zahnle et al. 2008; Nava-Sedeño et al. 2016; Wang et al. 2016; Catling et al. 2018). One relevant example is the planet Mars (Weiss et al. 2000; Sholes et al. 2018), which possesses an oxidized but $O_2$-poor atmosphere with no surface ocean, which is chemically and mechanistically different than the reducing Archean and oxygen-rich modern Earth scenarios we studied here. Furthermore, spectrally detectable CO without either significant $CH_4$ or $H_2O$ on an exoplanet could be suggestive of a limited or nonexistent biosphere if $CH_4$ lifetimes are expected to be high, and if orbiting an M star these factors may indicate an atmosphere conducive to generating abiotic $O_2$ via $CO_2$ photolysis (Gao et al. 2015; Harman et al. 2015; Schwieterman et al. 2016). However, we have shown that detectable CO may not always rule out the presence of life and may under some circumstances be compatible with—if not diagnostic of—a robust and productive biosphere.

It is important to note that the productivity of Earth's modern biosphere is not limited by photons or available energy, but instead by nutrients such as phosphorus, fixed nitrogen, or dissolved iron (Moore et al. 2013; Bristow et al. 2017). The ocean chemistry of Earth has changed over time and along with it trace metal limitations would have impacted the productivity of certain metabolisms. For example, it has been proposed that low copper availability in a euxinic Proterozoic ocean could have limited denitrification (reduction of $N_2O$ to $N_2$; Buick 2007). CO-consuming acetogens specifically require nickel as an essential component of carbon monoxide dehydrogenase (Dobbek et al. 2001). Indeed, nickel availability in the ocean is thought to have varied substantially over Earth's history, potentially inducing "nickel famines" with impacts on the productivity of methanogens and (presumably) acetogens as well (Konhauser et al. 2009, 2015). The possibility of nutrient limitation, coupled with evolutionary contingencies and ecosphere-atmosphere dynamics, should thus also constrain the applicability of antibiosignatures. Future studies examining the relationship between surface nutrient recycling—as influenced by weathering, oceanographic, and tectonic controls, among other factors—and exoplanet observables may allow us to place constraints on biospheric productivity and may potentially provide important context for evaluating putative antibiosignatures.

This study and other recent efforts suggest that the search for life elsewhere may yield ambiguous results because remote biosignatures will vary depending on planetary context or may not exist at all (Cockell 2014b; Deming & Seager 2017; Reinhard et al. 2017). Given the richness of potential outcomes, it is essential that we continue to develop frameworks and rubrics for identifying life (or ruling it out; Catling et al. 2018; Walker et al. 2018). Such research may be further bolstered by novel approaches such as cataloging potential alternative biosignature gases (Seager & Bains 2015; Seager et al. 2016), searching for biogenic seasonality (Olson et al. 2018a), advanced methods for calculating atmosphere-surface disequilibria (Krissansen-Totton et al. 2016, 2018b), and carefully considering the spectral capabilities of future space-based biosignature survey telescopes to ensure broad capabilities (Fujii et al. 2018; Kiang et al. 2018; Schwieterman et al. 2018a). Here, we have attempted to advance this cause by showing that measurements of appreciable CO in distant atmospheres may not always be evidence against the presence of life—thus elevating the importance of quantifying the environmental parameters and geologic boundary conditions for which this may be the case.

This work was also supported by the NASA Astrobiology Institute Alternative Earths team under Cooperative Agreement Number NNA15BB03A and the Virtual Planetary Laboratory (VPL) under Cooperative Agreement Number NNA13AA93A. The VPL is also supported by the NASA Astrobiology Program under grant number 80NSSC18K0829. E.W.S. is additionally grateful for support from the NASA Postdoctoral Program, administered by the Universities Space Research Association. S.L.O. acknowledges support from the T.C. Chamberlin postdoctoral fellowship in the Department of Geophysical Sciences at the University of Chicago. C.E.H gratefully acknowledges research support for the ROCKE-3D team through NASA's Nexus for Exoplanet System Science (NExSS), via solicitation NNH13ZDA017C. We thank Joshua Krissansen-Totton and the anonymous referee for helpful comments that improved our paper.

## ORCID iDs

Edward W. Schwieterman  https://orcid.org/0000-0002-2949-2163
Stephanie L. Olson  https://orcid.org/0000-0002-3249-6739
Chester E. Harman  https://orcid.org/0000-0003-2281-1990

## References

Abe, Y., Abe-Ouchi, A., Sleep, N. H., et al. 2011, AsBio, 11, 443
Anglada-Escudé, G., Amado, P. J., Barnes, J., et al. 2016, Natur, 536, 437
Arney, G., Domagal-Goldman, S. D., Meadows, V. S., et al. 2016, AsBio, 16, 873
Arney, G. N., Meadows, V. S., Domagal-Goldman, S. S. D., et al. 2017, ApJ, 836, 49
Bean, J. L., Abbot, D. S., & Kempton, E. M.-R. 2017, ApJ, 841, L24
Blomquist, B. W., Fairall, C. W., Huebert, B. J., et al. 2012, AMT, 5, 3069
Bolcar, M. R., Aloezos, S., Crooke, J., et al. 2017, Proc. SPIE, 10398, 9
Bristow, L. A., Mohr, W., Ahmerkamp, S., et al. 2017, Curr. Biol., 27, R474
Buick, R. 2007, Geobiology, 5, 97
Camacho, A., Walter, X. A., Picazo, A., et al. 2017, Front. Microbiol., 08, 323
Catling, D. C., Krissansen-Totton, J., Kiang, N. Y., et al. 2018, AsBio, 18, 709
Cockell, C. S. 2011, Trends Ecol. Evol., 26, 73
Cockell, C. S. 2014a, IJAsB, 13, 158
Cockell, C. S. 2014b, RSPTA, 372, 20130082
Cockell, C. S., Balme, M., Bridges, J. C., et al. 2012, Icar, 217, 184
Cockell, C. S., Bush, T., Bryce, C., et al. 2016, AsBio, 16, 89
Conrad, R., Seiler, W., Bunse, G., et al. 1982, JGR, 87, 8839
Conte, L., Szopa, S., Séférian, R., et al. 2018, BGD, 18, 881
Crisp, D. 1997, GeoRL, 24, 571
Deming, D., Seager, S., Winn, J., et al. 2009, PASP, 121, 952
Deming, L. D., & Seager, S. 2017, JGRE, 122, 53
Dittmann, J. A., Irwin, J. M., Charbonneau, D., et al. 2017, Natur, 544, 333
Dlugokencky, E. J., Nisbet, E. G., Fisher, R., et al. 2011, RSPTA, 369, 2058
Dobbek, H., Svetlitchnyi, V., Gremer, L., et al. 2001, Sci, 293, 1281
Domagal-Goldman, S. D., Segura, A., Claire, M. W., et al. 2014, ApJ, 792, 90
Fujii, Y., Angerhausen, D., Deitrick, R., et al. 2018, AsBio, 18, 739
Gao, P., Hu, R., Robinson, T. D., et al. 2015, ApJ, 806, 249
Gillon, M., Triaud, A. H. M. J., Demory, B.-O., et al. 2017, Natur, 542, 456
Greene, T. P., Line, M. R., Montero, C., et al. 2016, ApJ, 817, 17
Grenfell, J. L., Stracke, B., von Paris, P., et al. 2007, P&SS, 55, 661
Harman, C. E., & Domagal-Goldman, S. 2018, in Handbook of Exoplanets, ed. H. Deeg & J. Belmonte (Cham: Springer International), 71
Harman, C. E., Felton, R., Hu, R., et al. 2018, ApJ, 866, 56
Harman, C. E., Schwieterman, E. W., Schottelkotte, J. C., et al. 2015, ApJ, 812, 137
Kaltenegger, L. 2017, ARA&A, 55, 433
Kane, S. R., Hill, M. L., Kasting, J. F., et al. 2016, ApJ, 830, 1
Kasting, J. F., Liu, S. C., & Donahue, T. M. 1979, JGR, 84, 3097
Kharecha, P., Kasting, J., & Siefert, J. 2005, Geobiology, 3, 53
Kiang, N. Y., Domagal-Goldman, S., Parenteau, M. N., et al. 2018, AsBio, 18, 619
Konhauser, K. O., Pecoits, E., Lalonde, S. V., et al. 2009, Natur, 458, 750
Konhauser, K. O., Robbins, L. J., Pecoits, E., et al. 2015, AsBio, 15, 804
Krissansen-Totton, J., Bergsman, D. S., & Catling, D. C. 2016, AsBio, 16, 39
Krissansen-Totton, J., Garland, R., Irwin, P., et al. 2018a, AJ, 156, 114
Krissansen-Totton, J., Olson, S., & Catling, D. C. 2018b, SciA, 4, eaao5747
Lincowski, A. P., Meadows, V. S., Crisp, D., et al. 2018, ApJ, 867, 76
Luger, R., Sestovic, M., Kruse, E., et al. 2017, NatAs, 1, 0129
Martin, S. R., Rud, M., Mawet, D., et al. 2018, Proc. SPIE, 10698, 106980T
Meadows, V. S. 2017, AsBio, 17, 1022
Meadows, V. S., Arney, G. N., Schwieterman, E. W., et al. 2018a, AsBio, 18, 133
Meadows, V. S., & Crisp, D. 1996, JGR, 101, 4595
Meadows, V. S., Reinhard, C. T., Arney, G. N., et al. 2018b, AsBio, 18, 630
Moore, C. M., Mills, M. M., Arrigo, K. R., et al. 2013, NatGe, 6, 701
Nava-Sedeño, J. M., Ortiz-Cervantes, A., Segura, A., et al. 2016, AsBio, 16, 744
Neveu, M., Hays, L. E., Voytek, M. A., et al. 2018, AsBio, 18, 1375
Olson, S. L., Schwieterman, E. W., Reinhard, C. T., et al. 2018a, ApJL, 858, L14
Olson, S. L., Schwieterman, E. W., Reinhard, C. T., et al. 2018b, in Handbook of Exoplanets, ed. H. Deeg & J. Belmont (Cham: Springer International), 189
Ozaki, K., Tajika, E., Hong, P. K., et al. 2018, NatGe, 11, 55
Pavlov, A. A., Brown, L. L., & Kasting, J. F. 2001, JGRE, 106, 23267
Ragsdale, S. W. 2004, Crit. Rev. Biochem. Mol. Biol., 39, 165
Raiswell, R. 2006, J. Geochem. Explor., 88, 436
Reinhard, C. T., Olson, S. L., Schwieterman, E. W., et al. 2017, AsBio, 17, 287
Ricker, G. R., Winn, J. N., Vanderspek, R., et al. 2014, Proc. SPIE, 9143, 914320
Roberge, A., & Moustakas, L. A. 2018, NatAs, 2, 605
Robinson, T. D. 2017, ApJ, 836, 236
Robinson, T. D., Meadows, V. S., Crisp, D., et al. 2011, AsBio, 11, 393
Rugheimer, S., Kaltenegger, L., Segura, A., et al. 2015, ApJ, 809, 57
Rugheimer, S., Kaltenegger, L., Zsom, A., et al. 2013, AsBio, 13, 251
Schwieterman, E., Reinhard, C., Olson, S., et al. 2018a, arXiv:1801.02744
Schwieterman, E. W., Kiang, N. Y., Parenteau, M. N., et al. 2018b, AsBio, 18, 663
Schwieterman, E. W., Meadows, V. S., Domagal-Goldman, S. D., et al. 2016, ApJ, 819, L13
Seager, S., & Bains, W. 2015, SciA, 1, e1500047
Seager, S., Bains, W., & Petkowski, J. J. 2016, AsBio, 16, 465
Seager, S., Schrenk, M., & Bains, W. 2012, AsBio, 12, 61
Segura, A., Krelove, K., Kasting, J. F., et al. 2003, AsBio, 3, 689
Segura, A. A., Kasting, J. F., Meadows, V., et al. 2005, AsBio, 5, 706
Sholes, S. F., Krissansen-Totton, J., & Catling, D. C. 2018, AsBio, in press (arXiv:1811.08501)
Stevenson, K. B., Lewis, N. K., Bean, J. L., et al. 2016, PASP, 128, 094401
Sullivan, P. W., Winn, J. N., Berta-Thompson, Z. K., et al. 2015, ApJ, 809, 77
Walker, S. I., Bains, W., Cronin, L., et al. 2018, AsBio, 18, 779
Wang, Y., Tian, F., Li, T., et al. 2016, Icar, 266, 15
Weiss, B. P., Yung, Y. L., & Nealson, K. H. 2000, PNAS, 97, 1395
Zahnle, K., Haberle, R. M., Catling, D. C., et al. 2008, JGR, 113, E11004
Zhong, Q., Huang, Y., Shen, H., et al. 2017, Environ. Sci. Pollut. Res., 24, 864
Zsom, A., Seager, S., de Wit, J., et al. 2013, ApJ, 778, 109